# The Physics of Life: one molecule at a time

Mark C. Leake[1, 2] [*]

[1]Clarendon Laboratory, Dept of Physics, Parks Road, Oxford University, Oxford OX1 3PU, UK, [2]Dept of Biochemistry South Parks Road Oxford, OX1 3QU, UK.

[*]Correspondence: m.leake1@physics.ox.ac.uk

**ABSTRACT**

The esteemed physicist Erwin Schrödinger, whose name is associated with the most notorious equation of quantum mechanics, also wrote a brief essay entitled "What is Life?", asking: "How can the events in space and time which take place within the spatial boundary of a living organism be accounted for by physics and chemistry?" The 60+ years following this seminal work have seen enormous developments in our understanding of biology on the molecular scale, physics playing a key role in solving many central problems through the development and application of new physical science techniques, biophysical analysis and rigorous intellectual insight. The early days of single molecule biophysics research was centred around molecular motors and biopolymers, largely divorced from a real physiological context. The new generation of single molecule bioscience investigations has much greater scope, involving robust methods for understanding molecular level details of the most fundamental biological processes in far more realistic, and technically challenging, physiological contexts, emerging into a new field of "single molecule cellular biophysics". Here, I outline how this new field has evolved, discuss the key active areas of current research, and speculate on where this may all lead in the near future.

**1. INTRODUCTION**

Richard Feynman, celebrated physicist and bongo-drum enthusiast, gave a lecture in 1959 viewed by nanotechnologists of the future as a prophecy imagining perfectly their own field. The title was "There's plenty of room at the bottom", and it discussed a potential future to control and manipulate machines and store information and on a length scale tens of thousands times smaller than that of the everyday "macroscopic" world [1]. It was a clarion call to engineers and scientists to establish a new discipline, later coined *nanotechnology* [2]. Feynmann alluded to this small scale as relevant to that of biological systems, and how cells could function at this scale to perform "all kinds of marvelous things". We now know that this fundamental minimal unit is the single biological molecule. It's not to say that atoms comprising these molecules do not matter, nor sub-atomic particles that make up the individual atoms, nor smaller still the quarks of which the sub-atomic particles are composed. The point is, in general, we do not need to refer to length scales smaller than single molecules to understand most biological processes.

Technological developments in experimental biological physics have been the primary driving force in establishing the field of single molecule biophysics, and even though the discipline in its modern form is only a human generation in age it is clear that at the often prickly interfaces between the physical and the life sciences, and at scale of the single biological molecule, many of the most fundamental questions concerning cellular systems are being addressed. This field is evolving into a new discipline of *single molecule cellular biophysics* [3]. It is manifest not only in investigations at the single molecule level using live cells as the test system, i.e. *in vivo* single molecule studies, but also by some highly ingenious single molecule studies *in vitro* that, although divorced from the native physiological context, have a very high level of complexity either in the make up of the experimental components studied or in the combinatorial single molecule biophysics methods used, which greatly enhance the physiological relevance of the data obtained.

## 2. THE ESTABLISHMENT OF SINGLE MOLECULE BIOPHYSICS
### (a) *Why bother with single molecules?*

An experimental method which utilises *single molecule biophysics* gives us information on the position of a biomolecule in space at a given time or will allow the control and/or measurement of forces exerted by/on that molecule [4], or sometimes both. However, these approaches, despite being established for over two decades in dedicated scientific research laboratories around the world, are still technically challenging since they operate in a regime dominated by stochastic thermal fluctuations of water solvent molecules whose characteristic energy scale, that of $k_BT$ where $k_B$ is Boltzmann's constant and $T$ the absolute temperature measured in Kelvin, is comparable to energy transitions involved in molecular processes in biology. Forces are characterized by the piconewton (pN) scale, and the length scale of molecules and complexes is of the order of a few nanometres (nm), two orders of magnitude smaller than the wavelength of visible light (figure 1).

Why should we wish to perform such experiments which, as a rule, require measurements of tiny signals in environments of significant noise, in all but rare cases suffering from poor yields and, traditionally, being not remotely "high-throughput"? There already exist many robust bulk ensemble average biophysical methods which illuminate several aspects of structure and function of cellular systems using well-characterized experimental apparatus [5, 6], with an effect of averaging over copious molecular events, typically resulting in low measurement noise.

The principal reason for using novel physical methods and analyses for studying biological processes at the level of single molecules is the prevalence of *molecular heterogeneity.* One might suppose that the mean average property of ~$10^{19}$ molecules (roughly the number of molecules in 1 μl of water, equivalent to $1/(18 \times 1000)^{th}$ of a mole), as is the case for most bulk ensemble average techniques, is an adequate representation of the properties of any given single molecule. In some exceptional biological systems this is true, however, in general this is not the case. This is because single biological molecules usually exist in multiple states, intrinsically related to their biological functions. A state here is a measure of the energy locked in to that molecule. For example, there are many molecules which exist in multiple spatial conformations, such as *molecular motors*, with each conformation having a characteristic energy state.

Although there may be a single conformation which is more stable than the others for these tiny molecular machines, several shorter-lived conformations still exist which are utilized in different stages of motion and force generation. The *mean* conformation would look something close to the most stable of these many different conformations, but this single average parameter does not tell us a great deal about the behaviour of the other shorter-lived but essential states. Bulk ensemble average analysis, irrespective of what experimental property is measured, can not probe multiple states in a heterogeneous molecular system.

Also, *temporal fluctuations* in the molecules from a population result in broadening the distribution of a measured parameter from a bulk ensemble experiment which can be difficult to interpret physiologically. These thermal fluctuations are driven by *collisions* from the surrounding water molecules (~$10^9$ per second - biological molecules are often described as existing in a *thermal bath*) which can drive biological molecules into different states. In an ensemble experiment this may broaden the measured value, making reliable inference difficult. In single molecule measurements these states can often be probed individually.

Furthermore, there is a danger of lack of synchronicity in ensemble experiments. The issue here is that different molecules within a large population may be doing different things at different times, molecules may for example be in different conformations at a given time, so the average snapshot from the large population encapsulates all such temporal fluctuations resulting in a broadening of the distribution of any molecular parameter being investigated. The root cause of molecular asynchrony is that in most ensemble experiments the population is in steady-state, that is the rate of change between forward and reverse molecular states is identical. If the system is briefly taken out of equilibrium then transient molecular synchrony can be obtained, such as by forcing all molecules into just a single state, however this by definition is a transient effect so practical measurements are likely to be short-lived and technically challenging. These molecular-synchronizing methods include chemical and temperature jumps such as in stopped-flow reactions, electric and light field methods to align molecules, as well as freezing a population or causing it to form regular crystals. A danger with such approaches is that the normal physiological function may be different. Some biological tissues, for example cell membranes and muscles, are naturally ordered on a bulk scale and so these have historically generated the most physiologically relevant ensemble data.

The real strength of single molecule biophysics experiments is that these sub-populations of molecular states can be investigated. The importance to biology is that this multiple-state heterogeneity is actually an essential characteristic of the normal functioning of molecular machines; there is a fundamental instability in these molecules which allows them to switch between multiple states as part of their underlying physiological function.

A final point to note is that, although there is a wide range in concentration of biological molecules inside living cells, the actual number of molecules that are directly involved in any given biological process at any one time is generally low. Biological processes at this level can therefore be said to occur under minimal stoichiometry conditions in which just a few stochastic molecular events become important. In fact, it can often be these rarer, single molecule events that may be the most significant to cellular processes, and so it becomes all the more important to investigate life at the level of single molecules, and many approaches developed from the physical sciences have now been established focussed upon using single molecule biophysics techniques to address fundamental biological questions [7].

**(b)** *The first generation of single molecule biophysics investigations*
Single molecule biophysics is still a youthful field, in the context of the traditional "core" sciences. The first definitive single biological molecule investigations used pioneering electron microscopy techniques to produce metallic shadow replicas of large, filamentous molecules including DNA and a variety proteins [8], using fixed samples in a vacuum. Single particle detection began in non-biological samples, involving trapping single elementary particles in a gaseous-phase in the form of a single electron [9], and later as a single atomic ion [10].

The first single molecule biophysics investigation in which the surrounding medium included that one compound essential to all known forms of life, namely water, came with the fluorescence imaging in the lab of Boris Rotman in 1961 with the detection of single molecules of the enzyme *β-galactosidase* by chemically modifying one of its substrates to make it fluorescent, and observing the emergence of these molecules during the enzyme-catalysed reaction inside microscopic droplets [11] - although the sensitivity of detection at that time was not sufficiently high to monitor single fluorescent molecules directly, this particular assay utilised the fact that a single molecule of the *β-galactosidase* enzyme could generate several thousand substrate molecules which could be detected and thereby indicate the presence of a single enzyme. Comparable observations were made lab of

Thomas Hirshfeld over a decade later in aqueous solution without the need for microdroplets using the organic dye *fluorescein*, similar in structure to the fluorogenic component in the 1961 Rotman study, attached via antibodies to single *globulin* protein molecules, each with 80-100 individual fluorescein molecules bound [12]. The decade that followed involved marked developments in measurement sensitivity, including fluorescence detection of single molecules of a liquid-phase solution of the protein *phycoerythrin* labelled with ~25 molecules of the orange organic dye *rhodamine* [13], as well as parallel developments in the detection of single molecules in solids using *optical absorption* of a non-biological sample [14].

      The seminal single molecule biophysics work that came in the subsequent decade involved *in vitro* studies, experiments done, in effect, in the test tube. In the first instance, these investigations were driven by developments in a newly established technique of optical trapping, also known as laser or optical tweezers. The ability to trap particles using laser radiation pressure was reported by Arthur Ashkin, forefather of optical trapping, as early as 1970 [15], though the modern form which results in a net optical force on refractile/dielectric particles of higher refractive index than the surrounding medium roughly towards the intensity maximum of a focussed laser (figure 2*a-c*), was developed in the early 1980s by Ashkin and co-workers [16], and these optical force-transduction devices have since been applied with great diversity to study single molecule biophysics [17, 18].

      Arguably, the key pioneering biophysical investigation involving optical trapping used only a relatively weak optical trap in combination with a very sensitive sub-nm-precise detection technique called back focal plane interferometry [19], with micron-sized beads conjugated to molecules of the motor protein kinesin to monitor the displacement of single kinesin motors on a microtubule filament track, which indicated quantized stepping of each motor of a few nm consistent with the structural periodicity of kinesin binding sites on the microtubule [20]. This was followed by a study on another molecular motor of a type of myosin protein which was implicated in the generation of force during muscle contraction in its interaction with F-actin filaments [21].This investigation utilised two independent optical traps to tether a single filament and lower it onto a third, surface-immobilized, bead which

had been functionalized with the "motor-active" part of the myosin molecule. This was the first study to clearly measure both the quantized nature of displacement and force of a single molecular motor to nm/pN precision.

Biopolymer molecules were also the source of seminal single molecule biophysics investigations, using optical trapping to measure the mechanical molecular properties by stretching molecules and observing how the forces that developed changed with end-to-end displacement. These were applied to both single and double-stranded DNA [22] and RNA [23] nucleic acids (the latter study also investigating folding/unfolding transitions in the model RNA hairpin structural motif), as well as large modular proteins made up of repeating motifs of either the immunoglobulin or fibronectin family including many proteins related to the class of giant muscle proteins known as titins [24, 25, 26, 27].

A complementary technique of AFM force spectroscopy also emerged at around the same time. Surface probe techniques originated through the seminal work of Gerd Binning using the scanning tunnelling microscope (STM) [28] that measured *electron tunnelling* between a sample surface and micron-sized probe tip (a quantum mechanical effect whose probability depended exponentially on the tunnelling distance involved) as a measure of the surface topography. This developed into atomic force microscopy (AFM) [29], in which a similar probe tip, typically composed of silicon nitride, detects primarily *Van der Waals* forces from a sample surface, allowing imaging of surface topography to sub-nm precision. AFM force spectroscopy instead of imaging the surface uses a probe tip as a fishing-rod to clasp ends of molecules bound to gold-coated surface, and subsequently stretch them in retracting the tip away from the surface. This approach was used on modular protein constructs of titin to demonstrate forced unfolding of individual immunoglobulin modules. In doing so, this seminal paper showed evidence for a single molecule "signature" - a physical measurement indicating that there really is a single molecule under investigation, as opposed to multiples or noise, and in the case of AFM force spectroscopy this signature was a characteristic "sawtooth" pattern of the molecular force-extension trace that indicated dramatic changes in molecular extension of ~20-30 nm whenever one of the immunoglobulin modules made a forced transition from folded to unfolded conformations [30].

Developments in optical imaging, most importantly fluorescence microscopy, had an enormous impact on pushing single molecule biophysics forward. These have included molecular interaction methods using single molecule Förster resonance energy transfer (smFRET) in which energy can be transferred non-radiatively between differently coloured donor and acceptor dye molecules, each designed to be attached to biological structures which transiently interact as part of their biological function. FRET occurs provided there is suitable spectral overlap between the emission and absorption spectra, and the two molecules are both oriented appropriately and within less than ~10 nm of each other. The first clear report of smFRET measurements involved monitoring single molecule assembly of the DNA double helix [31].

Fluorescence imaging was also applied to monitor rotation of single molecules of the rotary motor F1-ATPase by attachment of a rhodamine-tagged fluorescent filament of F-actin conjugated to the F1-ATPase rotor subunit, which demonstrated clear rotation of this vital biological machine responsible for the generation of the universal cellular fuel ATP, but also showed the motion occurs in quantized angular units mirroring the symmetry of the enzyme's atomic structure [32].

In another pioneering study, single molecule fluorescent dye imaging was used to monitor the movement of tagged myosin molecules to show that they travelled along F-actin tracks in a *hand-over-hand* mechanism. This was the first study to show *unconstrained* walking of a single molecular motor, using nm-precise localization in the form of Gaussian fitting of the "point spread function" image of each single fluorescent dye molecule, which the investigators denoted as *fluorescence imaging with one nanometre accuracy*, or FIONA [33].

A seminal *in vitro* study which links to several key *in vivo* investigations involved the application of high-speed millisecond fluorescence imaging to monitor real-time diffusion of single lipid molecules labelled with an organic dye, expressed in an artificial lipid bilayer [34], thus acting as a mimic for real cell membranes. Here, investigators could track single molecules with an accuracy better than the optical resolution limit (~200-300 nm) using a method which estimated the centre of the fuzzy diffraction-limited intensity image of single dye molecules to within a few tens of nm precision by using Gaussian fitting to the raw images (a method that was originally applied almost a decade earlier to determine the centre position of 190 nm diameter kinesin-coated beads conjugated to microtubules from non-fluorescence brightfield differential interference contrast (DIC) images to within 1-2 nm precision [35]).

## 3. THE "GOLDEN AGE" – THE EMERGENCE OF SINGLE MOLECULE "CELLULAR" BIOPHYSICS

**(a)** *Approaches that investigate living, functional cells*

With so much exemplary single molecule biophysics research performed in the test tube, a question which should be addressed is: why do we care about studying molecular details in live-cell, or near live-cell, environments? Test tube environments are significantly more controllable, less contaminated and come associated with less measurement noise. The best answer is that cells are not test tubes. A test tube experiment is a much reduced version of the native biology containing only components which we think/hope are important. We now know definitively that even the simplest cells are not just bags of chemicals, but rather have localized processes in both space and time. Also, the effective numbers of molecules involved in many cellular processes are often low, sometimes just a few per cell, and these minimal stoichiometry conditions are not easy to reproduce in the test tube without incurring a significant reduction in physiological efficiency.

Single molecule biophysics investigations *in vivo* are, however, technically very difficult. Here, fluorescence microscopy is an invaluable biophysical tool. It results in exceptionally high signal-to-noise ratios for determining the localization of molecules tagged with a fluorescent dye but does so in a way that is relatively non-invasive compared to other single molecule biophysics methods. This minimal perturbation to native physiology makes it a probe of choice in single molecule biophysics studies in the living cell. Many of the improvements in our ability to detect single molecules have been driven by developments in the technology that allows photons to be efficiently collected from molecular report probes, several of which are fluorescent, including both "point" detectors such as the photomultiplier tube (PMT) to pixel arrays of the next generation high quantum-efficiency cameras called electron multiply charge-coupled devices (EMCCDs), and these comparative technologies are reviewed in this Theme Issue [36].

It was only as recently as the year 2000 that the first definitive single molecule biophysics investigation involving a living sample was performed - by Sako and others [37] in which the investigators performed single molecule live-cell imaging on the cell membrane, here the high-contrast imaging technique of total internal reflection fluorescence microscopy (figure 3*a*), or TIRF [38], monitoring fluorescently-labelled EGF ligands binding to membrane receptor, and by Byassee and others [39] in which the researchers performed single molecule live-cell imaging inside the centre of a cell using confocal microscopy to monitor fluorescently-labelled transferrin molecules undergoing endocytosis.

Significant developments have been made over the past decade in the field of *live-cell super-resolution imaging* [40],the ability to perform optical imaging *in vivo* at a spatial resolution better than that predicted from the Abbe optical resolution limit of $\sim 0.61\lambda/NA$, where $\lambda$ is the detected wavelength for imaging and *NA* is the numerical aperture of the imaging system (typically set by the objective lens of the optical microscope of $\sim$1.2-1.5), in particular an ability to monitor functional molecular complexes with such precision [41, 42]. There are several reviews that the reader can seek to discover the state-of-the-art in regards to various super-resolution technologies, however in this Theme Issue, super-resolution methods are reviewed in the context of a relatively new and highly promising technique called optical lock-in detection (OLID) which permits dramatic improvements to imaging contrast in native cellular imaging, far in excess of other competing super-resolution methods [43].

Recent developments in cellular single molecule fluorescence imaging have include the ability to definitively count molecules that are involved in functional biological processes integrated in the cell membranes of live cells, for example to quantify multiple protein subunit components in relatively large molecular machines such as the bacterial flagellar motor [44, 45] or single ion channels [46], and to combine counting with tracking of relatively mobile components around different spatial locations in the cell, such as molecular machines involves in protein translocation [47] and ATP fuel generation via oxidative phosphorylation, or OXPHOS [48, 49]. The state-of-art of our ability to image molecular components in cell membranes has led to substantial improvements to our understanding of their complex architecture, reviewed in two articles in this Theme Issue for model bacterial systems [50] as well focussing on putative zones of molecular confinement in the membrane, commonly referred to as *lipid rafts* [51]. By modifying the modes of fluorescence illumination, for example using narrow-field [34] or slimfield imaging [52], it has been possible to increase the excitation intensity in the vicinity of single cells to allow millisecond single molecule imaging. This has permitted visualization of native components normally expressed in the cytoplasm of cells whose viscosity is 100-1,000 times smaller than that of the cell membrane and so would be expected to diffuse at a faster rate by this same factor, allowing observation of gene expression bursts [53], regulation of transcription factors [54] and quantification of functional *replisome* components used in bacterial DNA replication machines [55].

Despite the central importance of fluorescence methods for single molecule cellular imaging there are also non-fluorescence detection techniques which can generate highly precise. For example, scanning probe microscopy (SPM) techniques. These cover a range of experimental approaches allowing *topographical detail* from the surface of a sample to be obtained by laterally scanning a probe across the surface. There are more than 20 different types of SPM methods currently developed which measure a variety of physical parameters as the probe is placed in proximity to a sample surface, and the most popular to date has been AFM (figure 3*b*). In this Theme Issue, Klenerman et al [56] reviews SPM techniques in the context of singe molecule precise imaging on the topographical details of live cells, namely probe-accessible features present on the cell membrane, and discusses in depth a relatively novel SPM approach of scanning ion conductance microscopy, or SICM (figure 3*c*).

Another non-fluorescence technique which shows significant future potential for single molecule cellular biophysics is surface enhanced Raman scattering (SERS). Raman scattering is an inelastic process such that scattered photons from a sample have a marginally different frequency to those of the incident photons due primarily to vibrational energy transfer from the molecular orbitals in the sample, either resulting in a loss of energy from the photons (Stokes scattering) or, less commonly, a gain (anti-Stokes scattering). However, to detect the presence of a single molecule in a sample using Raman spectroscopy requires significant enhancement to the standard method used to acquire a scattering spectrum from a bulk, homogeneous sample. The most effective method utilises surface enhancement, which is reviewed in this Theme Issue [57], involving placing the sample in a colloidal substrate of gold or silver nanoparticles tens of nm in diameter. Photons from a laser will induce *surface plasmons* in the metallic particles, and in the vicinity of the surface the local electric field $E$ associated with the photons is enhanced by a factor $E^4$. The enhancement depends critically on the size/shape of the nanoparticles, but typically generates a better measurement sensitivity by a factor $\sim 10^{14}$, particularly effective if the molecule itself is conjugated to the nanoparticle surface. This enhancement can be sufficient to detect single biomolecules.

### (b) *In vitro methods of high complexity*

This is not to say that *in vitro* experiments are intrinsically bad and *in vivo* experiments are definitively good. Rather, they each provide *complementary* information.

*In vitro* experiments are detached from a true physiological setting, but the level of environmental control is high. *In vivo* experiments are more demanding technically and are subject both to greater experimental noise and intrinsic biological variation - being in a native physiological environment is appealing at one level but offers difficulty in interpretation since there is a potential lack of control over other biological processes not directly under study but which may influence the experimental results.

Next generation *in vitro* single molecule biophysics approaches are characterized by a much greater complexity than those involved in the early days of the field. In this Theme Issue, some of these often highly involved novel test tube approaches are discussed in Duzdevich and Greene [58], with a particular emphasis on a high-throughput single molecule biophysics method to investigate the binding of proteins to DNA, called *DNA curtains*.

One particular focus of recent *in vitro* single molecule experiments has been the FoF1-ATPase enzyme, a highly complex machine composed of two rotary molecular motors of the membrane-integrated Fo motor and the hydrophilic F1 motor, which are ultimately responsible for the generation of cellular ATP. In this Theme Issue, recent single molecule biophysics approaches to investigate this vital, ubiquitous enzyme are reviewed in Sielaff and Börsch [59], with novel confirmation that the mechanism of nanoscale stepping of the F1 component elucidated in a *thermophilic* enzyme at room temperature, in which molecular rotation has been fuelled by the hydrolysis of ATP in the opposite direction to that involved during ATP manufacture, is shared by the *mesophilic E. coli* F1 enzyme, suggesting that even in markedly different environments there are common modes of action to this ubiquitous, essential molecular machine (Bilyard et al [60]).

**(c)** ***Novel automated and bio-computational techniques***

Single molecule biophysics experiments are often plagued with noise, with the effective signal-to-noise ratio being sometimes barely in excess of 1 and generally less than 10. This constitutes an enormous analytical challenge to reliably detect a true signal and not erroneously measure noise. Molecular events are often manifest as some form of transient *step* signal in a noisy time-series, for example a motor protein might move via stepping along a molecular track. Thus, the challenge becomes one of reliable *step-detection* from noisy data. The aim is to assemble quantitative statistics of such step events in a fully objective, automated way.

*Edge-preserving filtration* of the raw, noisy data is often the first tool employed, which preserves distinct edge event in time-series, such as the simple *median filter*, or better still the *Chung-Kennedy filter* which consists of two adjacent running windows whose output is the mean from the window possessing the smallest variance [26, 27] - a step event may then be classed as "true" on the basis of the change in the mean and variance between the two windows being above some pre-agreed threshold.

A significant issue with step-detection from a data time-series is that detection is sensitive to the level of threshold set. An alternative approach where all steps in a series are expected to be of the same size is to convert the time-series into a *frequency-domain* using a Fast Fourier transform, and then detect the periodicity in the original trace by looking for a fundamental peak in the associated *power spectrum*, which has been used to good effect for the estimation of molecular stoichiometry using step-wise photobleaching of fluorescent proteins [44].

A recent improvement to objectifying single molecule biophysics data is in how the *distributions* of single molecule properties are rendered. Traditional approaches used *histograms*, however these are highly sensitive to histogram bin size and position. A more general, objective approach uses *kernel density estimation* (KDE) - data are *convolved* with a Gaussian whose width is the *measurement error* for that property in that particular experiment, and whose height is normalized so that the area under the Gaussian is precisely one (i.e. one detected event), used to good effect in studying single molecule architectures of the bacterial replisome [55].

Spatial dynamics of single molecules and complexes inside living cells is a feature of biological processes. However, due to the low signal-to-noise ratio involved in cellular imaging experiments, the analysis of the motions of molecular complexes is non-trivial. In this Theme Issue, Robson et al [61] describe a novel method implementing a well-known weapon in the statistician's armoury called *Bayesian inference* to robustly determine the underlying different modes of molecular diffusion relevant to live-cell imaging in both an objective and automated manner.

One of the biggest challenges to single molecule biophysics is the traditionally low-throughput nature of experiments. In this Theme Issue, Ullman et al [62] describe methods combining automated microfluidics and novel imaging/analysis to dramatically improve the high-throughput nature.

## 4. THE CONTRIBUTIONS IN THIS THEME ISSUE

This Theme Issue presents a series of articles from leaders in the field offering new insight into some of the latest developments of single molecule biophysics research which has now moved towards a far greater physiological relevance into the regime of addressing real, cellular questions. In summary, these articles include:

i. A comprehensive review of new approaches in photon detection technology essential to modern single molecule cellular biophysics research [36].

ii. Novel insights into super-resolution fluorescence imaging using the exceptionally high-contrast method of optical lock-in detection, OLID [43].

iii. An appraisal of the increasing use of model bacteria as experimental testbeds for addressing fundamental biological questions using single molecule techniques [50].

iv. A robust comparison of the single molecule biophysics methods which probe the nanoscale architectures of lipid microdomains in cell membranes [51].

v. A description of new, exciting single molecule surface probe technologies for living cells, including surface ion conductance microscopy, SICM [56].

vi. A discussion of promising new single molecule cellular biophysics probing techniques using surface enhanced Raman spectroscopy, SERS [57].

vii. A review of elegant, *in vitro* approaches to comb out single DNA tethers for investigating single molecule protein translocation [58].

viii. An exploration of the state-of-the-art in single molecule biophysics methodologies for experimentally probing the molecular means of ATP generation in cells [59].

ix. Novel, cutting-edge single molecule biophysics research showing how the rotary molecular motors used in ATP generation in cell species which experience markedly different physical environments share fundamental mechanistic features [60].

x. New research illustrating powerful new bio-computational approaches to characterize the underlying modes of molecular diffusion from live-cell single molecule imaging [61].

xi. A novel investigation demonstrating how single molecule experiments on live cells can be made substantially more high-throughput by utilising ingenious engineering developments in microfluidics and computational improvements to optical microscope automation [62].

## 5. THE OUTLOOK - BEYOND THE SINGLE MOLECULE AND THE SINGLE CELL

The development of single molecule cellular biophysics represents a coming-of-age of methods using physics to understand life at the molecular level. There is great potential to now apply these novel technologies into areas that may have a large future impact on society, including those of *bionanotechnology*, *systems* and *synthetic biology*, *fuel production* for commerical use and *single molecule biomedicine*.

As a scientific field, single molecule cellular biophysics is undergoing enormous expansion and is likely to be a key discipline in revealing underlying mechanistic features of biological processes in cells, with significant implications for the shape of both biophysical and biomedical research in the future. The industrial motivation to miniaturize synthetic bio-inspired devices is already starting to feedback into academic research laboratories in catalysing a general *down-sizing* approach for measurement apparatus.

There is a compelling need to push this area of physiologically relevant interfacial science forward significantly, and this can only be truly facilitated by future generations of life and physical scientists *talking* to each other. Folk from each side of the bioscience fence traditionally blend like oil and water, such immiscibility often stemming from unfortunately early academic choices that schoolchildren make. However, what is needed now is an appreciation that some of the most fundamental concepts in each discipline can be shared by both camps, once elements of unwieldy language and overly complex maths have been put aside.

The outlook for single molecule cellular biophysics is highly promising, but it is fundamentally driven by the enthusiasm of the talented researchers willing to take a punt and cross bridges into areas of science unknown.

Preparation of this article was supported by a Royal Society University Research Fellowship and EPSRC research grant (EP/G061009) to M.C.L. The help of the anonymous referees who commented on contributions to this Theme Issue is gratefully acknowledged, as is the diligent assistance and patience of Helen Eaton of the Royal Society during the publication process.

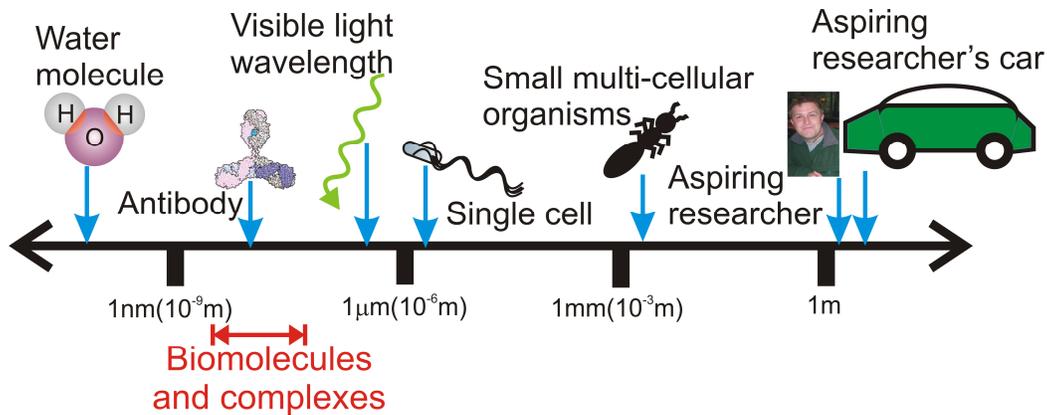

Figure 1. A schematic representation of the length scale of biological molecules and complexes in the context of larger macroscopic length scale entities.

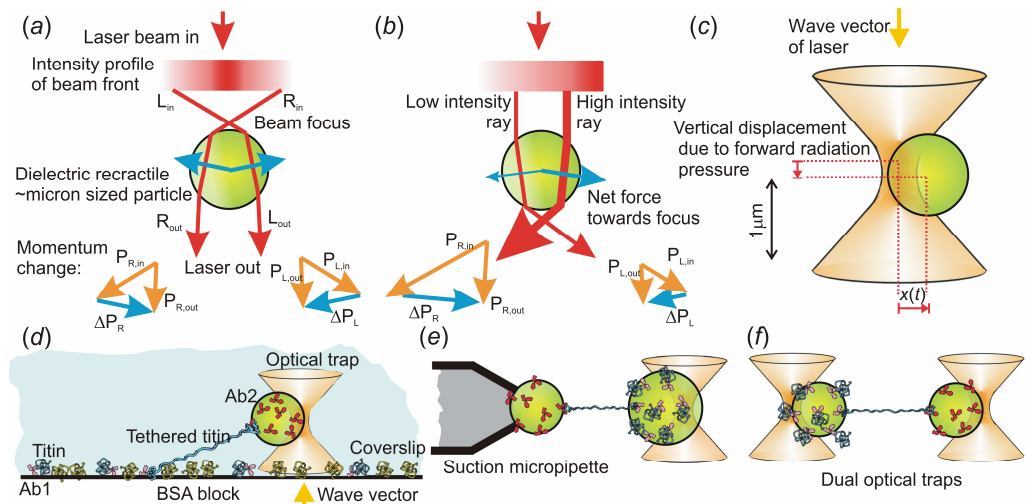

Figure 2. Optical trapping. (*a*) Ray-optic depiction of the trapping force for an optically trapped particle - a parallel Gaussian-profile laser beam is focussed and refracted by the trapped particle, such that equal and opposite changes in momentum on either side of the particle cancel out resulting in zero net force when the particle is roughly at the laser focus. But, (*b*) when the particle is laterally displaced from the focus the net momentum change experienced due to the reaction forces when refracted beams of light emerge from the particle are directed back towards the laser focus, illustrated by the momentum vector plots. (*c*) Displacement of a micron sized bead in an optical trap, the lateral trapping force is proportional to the lateral displacement $x(t)$ where time is time, also the forwarded radiation pressure pushes the bead a little away from the precise laser focus. (*d*) Single optical trap stretch of a titin molecule tethered to a microscope coverslip via antibodies Ab1 and Ab2 binding to opposite termini of the titin molecule. (*e*) Similar titin stretches using a suction micropipette combined with an optical trap and (*f*) dual optical traps.

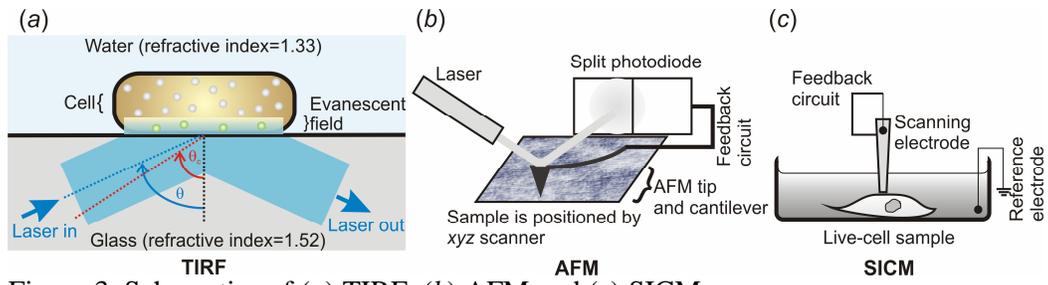
Figure 3. Schematics of (*a*) TIRF, (*b*) AFM and (*c*) SICM.